\documentstyle[12pt]{article}
\def\lappeq{\mathrel{\rlap{\raise.5ex\hbox{$<$}}
{\lower.5ex\hbox{$\sim$}}}}
\def\beq{\begin{equation}}
\def\eeq{\end{equation}}
\begin{document}
\begin{titlepage}
\pagestyle{empty}
\baselineskip=21pt
\rightline{Alberta Thy-19-99, CERN-TH/99-392}
\rightline{hep-ph/0001122}
\rightline{January 2000}
\vskip .05in
\begin{center}
{\large{\bf  Reheating and Supersymmetric Flat-Direction Baryogenesis}}
\end{center}
\vskip .05in
\begin{center}
{\bf Rouzbeh Allahverdi}, {\bf Bruce. A. Campbell} 

{\it Department of Physics, University of Alberta}

{\it  Edmonton, Alberta, Canada T6G 2J1}

and \\
{\bf John Ellis}

{\it Theory Division, CERN}

{\it CH-1211 Geneva 23, Switzerland}

\vskip .05in

\end{center}
\centerline{ {\bf Abstract} }
\baselineskip=18pt

\noindent

We re-examine Affleck-Dine baryo/leptogenesis from the oscillation of
condensates
along flat directions of the supersymmetric standard model, which
attained large vevs at the end of the inflationary epoch. The
key
observation is that superpotential interactions couple the flat
directions to other fields, which acquire masses induced by the flat-direction
vev that may be sufficiently small for them to be kinematically accessible to
inflaton decay. The resulting plasma of inflaton decay products then
may act on the flat directions via these superpotential Yukawa
couplings, inducing thermal masses and supersymmetry-breaking
$A$ terms. In such cases the flat
directions start their oscillations at an earlier time than usually
estimated. The oscillations are also terminated earlier, due to
evaporation of the flat direction condensate produced by its
interaction with the plasma of inflaton decay products.
In these cases we find that estimates for the resulting baryon/lepton
asymmetry of the universe are substantially altered. We identify
scenarios for the Yukawa couplings to the flat directions, and the
order and mass scale of higher-dimensional superpotential interactions that set the initial flat direction vev, that might lead to acceptable baryo/leptogenesis.

 \vskip .1in

\centerline{{Submitted For publication In: {\it Nuclear Physics B }}}

\end{titlepage}
\baselineskip=18pt
{\newcommand{\la}{\mbox{\raisebox{-.6ex}{$\stackrel{<}{\sim}$}}}
{\newcommand{\ga}{\mbox{\raisebox{-.6ex}{$\stackrel{>}{\sim}$}}}


\section{Introduction}

Initially, one of the most attractive features of Grand Unified
Theories
(GUTs) \cite{ross} was the prospect that they might provide an explanation \cite{sakh} for the
matter-antimatter asymmetry of the Universe, via their new interactions
that violate baryon and/or lepton number.  Subsequently, it has been
realized that, even in the Standard Model, at the non-perturbative
level
there are sphaleron interactions that violate both baryon and lepton
number. This discovery has given rise to new scenarios for
baryogenesis,
at the electroweak phase transition \cite{krs1} or via leptogenesis followed by
sphaleron reprocessing \cite{fy}. Supersymmetric extensions of the Standard
Model offer yet more scenarios for baryogenesis. For example, they
may facilitate electroweak baryogenesis by permitting a first-order
electroweak phase transition despite the constraints imposed by
LEP~\cite{LEPBG}.
There is also the possbility that they may contain perturbative
interactions that violate baryon and/or lepton number via
a breakdown of $R$ parity, which under certain circumstances \cite{cdeo} can induce baryogenesis.

However, perhaps the most attractive
mechanism offered by supersymmetry is that proposed by Affleck and
Dine,
according to which \cite{ad} a condensate of a combination of
squark and/or slepton fields may have formed during an inflationary epoch \cite{infl} in the early universe,
causing the vacuum to carry a large net baryon and/or lepton number,
which is then transferred to matter particles when the condensate
eventually decays. We recall
that the condensate forms along some flat direction of the
effective potential of the theory, which we take to be the Minimal
Supersymmetric extension of the Standard Model (MSSM) at low
energies. In the conventional approach to Affleck-Dine
baryogenesis, the condensate is essentially static until a
relatively late cosmological epoch, when it starts to
oscillate. In turn, the termination of the period of
oscillation has been calculated in terms of the magnitudes
of the soft supersymmetry-breaking terms present in the
effective potential, which become significant only at low
temperatures, and of the thermalization effects of inflaton decay \cite{eeno}.

The purpose of this paper is to re-examine this Affleck-Dine
mechanism by incorporating a more complete treatment of the
reheating of the universe after the inflationary epoch. We
argue that the flat directions are in general coupled to
other fields that are kinematically accessible to inflaton
decay. These fields therefore have non-trivial statistical
densities, and become thermalized. The couplings
of these densities to the
flat directions induce effective supersymmetry-breaking
masses and $A$ terms for the erstwhile flat fields.
As a result, the `flat' directions start oscillating earlier than
previously estimated. Subsequently, the oscillations also
terminate earlier, as the flat-direction condensate interacts with the
plasma of inflaton decay products and evaporates. The bottom line
is that previous estimates of the resulting baryon/lepton
asymmetry of the universe may be substantially altered, and
we estimate some orders of magnitude for different
representative parameter choices.


\section{Flat Directions}

The $D$-flat directions of the MSSM are classified by gauge-invariant
monomials in the fields of the theory.  These monomials have been classified in
\cite{gkm}, and, for directions which are also $F$-flat
for renormalizable standard model superpotential interactions, the
dimension
of the non-renormalizable term in
the superpotential which first lifts the respective $D$-flat direction has
also been
derived.
Hereafter, we consider only those $D$-flat directions which are not
lifted by renormalizable superpotential interactions.  These correspond
to 14 independent monomials, and each monomial represents a complex
$D$-flat direction: one vev magnitude and one phase (all fields in the
monomial have the same vev).  Since the monomials are gauge-invariant,
appropriate gauge transformations generated by non-diagonal
generators can be used to remove that part of the $D$-term contribution to
the potential which comes
from the non-diagonal generators.  Also, any
relative phase among the fields in the monomial can be rotated away by
those gauge transformations which are generated by diagonal generators.
There remains only one overall phase, i.e., the phase of the flat
direction,
which can be absorbed by redefinition of the scalar fields.  Note that the
$D$-term and $F$-term parts of the scalar potential are invariant under
such a redefinition (which is equivalent to a U(1) symmetry
transformation) while the soft-breaking terms and fermionic Yukawa
terms generally are not.  So we can always arrange the vev of the
fields in the monomial to be initially along the real axis.  We also
note that such a non-zero vev breaks spontaneously the
MSSM gauge group.

As an explicit example, consider the simplest case, which is the
${H}^{u}L$ flat direction.  If the $T_3= {1 \over 2}$ component of
${H}^{u}$ and the $T_3=-{1 \over 2}$ component of $L$ have the same vev,
then all the $D$ terms from both diagonal and non-diagonal generators of
the MSSM are zero.  The non-diagonal ones are identically zero and the
equality of the vev makes the diagonal ones zero as well.  These
vev's can then be chosen along the real axis as noted above.  There
are eight real degrees of freedom in the ${H}^{u}$ and $L$ doublets.
Two of them comprise the flat direction and another three are Goldstone
bosons eaten by the gauge fields of the spontaneously-broken
symmetries.  The remaining three are physical scalars which are coupled
to the flat direction, and are massive due to its vev.

Now that all fields in the monomial have the same vev and are real,
by an orthogonal transformation we can go to a new basis where there is
only one direction with a non-zero vev.  Let us label this direction
$\alpha$ and the orthogonal directions generically as $\phi$, therefore
${\alpha}_{R}\neq 0$ while ${\alpha}_{I} =
{\phi}_{R} = {\phi}_{I} = 0$. For the specific ${H}^{u}L$
example, these are the following combinations after the Goldstone bosons
are
absorbed by the Higgs mechanism:

\begin{eqnarray*}
\sqrt{2} {\alpha}_{R} &= &{({H}_{1})}_{R}+{({L}_{2})}_{R}\\
\sqrt{2} {\alpha}_{I} &=& {({H}_{1})}_{I}+{({L}_{2})}_{I}\\
\sqrt{2} {\phi}_{1} &=& {({H}_{1})}_{R}-{({L}_{2})}_{R}\\
\sqrt{2} {\phi}_{2} &=& {({H}_{2})}_{R}-{({L}_{1})}_{I}\\
\sqrt{2} {\phi}_{3} &=& {({H}_{2})}_{I}+{({L}_{1})}_{R}
\label{HuL}
\end{eqnarray*}
The $D$ terms from the ${T}_{3}$ and ${U(1)}_{Y}$ generators give terms
${g}^{2}{\alpha}^{2}{{\phi}_{1}}^{2}$ (up to numerical factors) in the
potential, whilst
those from ${T}_{1}$ and ${T}_{2}$ give 
${g}^{2}{\alpha}^{2}{{\phi}_{2}}^{2}$ 
and ${g}^{2}{\alpha}^{2}{{\phi}_{3}}^{2}$ 
terms (up to numerical factors).  It is a generic feature
that all fields entering in the flat direction monomial which are left
after the Higgs mechanism (except the linear combination which receives
the vev after diagonalization) have masses of order $g \alpha$ due to
their $D$-term couplings to the flat-direction vev.

We now consider supergravity effects, both in minimal models with
soft-breaking terms at the tree level, and in no-scale 
models~\cite{LN}, where such terms
are absent at tree-level but arise from quantum corrections \cite{mk}.  
The superpotential consists
of the tree-level MSSM terms and a series of non-renormalizable terms of
successively higher dimension, which are induced in the effective theory
by the dynamics of whatever is the underlying more fundamental theory.
Without imposing  $R$~parity (or any other symmetry) all gauge-invariant
terms of higher dimension would exist in the superpotential. We may,
however, also wish to impose $R$ parity on the higher-dimensional terms,
as we have done on the renormalizable interactions, to prevent
substantial $R$-parity violation being fed down from high scales by the
renormalization-group running of the soft mass terms \cite{aco1}. If we
assume that
$R$ parity is a discrete gauge symmetry of the theory, then it would be
respected by all gauge-invariant superpotential terms of arbitrary
dimension.  Relevant higher-dimensional superpotential terms which lift
the flat direction $\alpha$ are of the form:

\beq
W \supseteq {\lambda}_{n} {{\alpha}^{n} \over n{M}^{n-3}}
\eeq
where ${\lambda}_{n}$ is a number of order one and $M$ is a large mass
scale, e.g., the GUT or Planck scale.

During inflation, supersymmetry is strongly broken by the non-zero
energy of the vacuum.  In minimal models this is transferred to the
observable sector through the K\"ahler potential at tree level \cite{drt},
while
in no-scale models~\cite{LN} this happens at the one-loop level
\cite{gmo}.  Inflation then induces \cite{drt} the soft-breaking terms

\beq
-{C}_{I}{{H}_{I}}^{2}{|\alpha|}^{2} + a {\lambda}_{n}{H}_{I}
{{\alpha}^{n} \over n{M}^{n-3}} + h.c.
\eeq
where ${C}_{I}$ and $a$ are numbers depending on the sector in which the inflaton
lies, and ${H}_{I}$ is the Hubble constant during inflation.  We
shall assume here that ${C}_{I}$ is positive and not unnaturally small
\cite{drt,gmo}.  In the presence of the $A$ term, the potential
along the angular direction has the form $\cos{(n\theta +{\theta}_{a})}$,
where ${\theta}_{a}$ is the phase of $a$.  Due to its negative
mass-squared, the flat direction rolls down towards one of the discrete
minima at $n\theta +{\theta}_{a}=\pi$ and $|\alpha| = {({{C}_{I} \over
(n-1){\lambda}_{n}}
{H}_{I}{M}^{n-3})}^{{1 \over n-2}}$, and quickly settles at one of the
minima (${{C}_{I} \over (n-1){\lambda}_{n}} $ is $O(1)$).  Therefore, at the
end of inflation $\alpha$ can be at any of the above-mentioned
minima.

In the absence of thermal effects, $\alpha$ would track the
instantaneous minimum $|\alpha|  \sim {( H {M}^{n-3})}^{{1 \over
n-2}}$ from the end of inflation until the time when $H \simeq {m}_{{3
\over 2}}$, where $ {m}_{{3 \over 2}} \sim 1$~TeV is the low-energy
supersymmetry-breaking scale \cite{drt}. At $H \simeq {m}_{{3 \over 2}}$
the low-energy soft terms 

\beq
{{m}_{{3 \over 2}}}^{2} {|\alpha|}^{2} + A {\lambda}_{n} {m}_{{3 \over 2}} {{\alpha}^{n} \over
n{M}^{n-3}} + h.c.
\eeq
would take over, with the mass-squared of $\alpha$
becoming positive, and $\alpha$ would then start its
oscillations.  Also, the minima along the angular direction would then
move in a non-adiabatic way, due generally to different phases for $A$
and $a$.  As a
result, $\alpha$ starts its free oscillations around the origin with an
initial vev ${\alpha}_{osc.} \sim {({m}_{{3 \over 2}}{M}^{n-3})}^{{1
\over n-2}}$ and frequency ${m}_{{3 \over 2}}$ and, at the same time,
the torque exerted on it causes motion along the angular direction.  In
the case that the flat direction carries a baryon (lepton) number this
will lead to a baryon (lepton) asymmetry ${n}_{B}$ \cite{ad} given by
${n}_{B} = {\alpha}_{R} {\partial {{\alpha}_{I}}\over 
\partial t} - {\alpha}_{I} {\partial {{\alpha}_{R}}\over \partial t}$.  
At ${m}_{3 \over 2} t \gg 1$ the 
upper bound on ${n}_{B}$~\cite{ad} may be written 

\beq
{n}_{B} \sim {1 \over {{m}_{3 \over 2}}^{2} {t}^{2}}{{\alpha}_{osc.}}^{3}{({{\alpha}_{osc.} \over M})}^{n-3}
\eeq
which, after transition to a radiation-dominated universe, results in an
${{n}_{B} \over s}$ that remains constant as long as there is no further
entropy release.

As we will see in subsequent sections, thermal effects of inflaton
decay products with superpotential couplings to the flat direction can
fundamentally alter the dynamics of the flat direction oscillation, and
necessitate revision of the estimates for the resulting baryon/lepton
asymmetries produced.


\section{Flat-Direction Superpotential Couplings and Finite Temperature Effects}

As we have seen, the flat direction $\alpha$ has couplings of the form
${g}^{2}{\alpha}^{2}{\phi}^{2}$ to the fields $\phi$ which are in the
monomial that represents it.  Besides these $D$-term couplings, it also
has $F$-term couplings to other fields $\chi$ which are not present in
the monomial.  These come from renormalizable superpotential Yukawas,
and have the form
\footnote{We note that for $F$-flat directions of the renormalizable
piece of the superpotential, which are only lifted by
higher-dimensional nonrenormalizable terms, $\alpha$ cannot have such
superpotential couplings to $\phi$ fields which appear in the
monomial.}

\beq
W \supseteq h \alpha \chi \chi
\eeq
which results in a term ${h}^{2}{|\alpha|}^{2}{|\chi|}^{2}$ in the scalar
potential.
Again for illustration, consider the ${H}^{u}L$ flat direction:
${H}^{u}$ has Yukawa couplings to left-handed and right-handed
(s)quarks while $L$ has Yukawa couplings to ${H}^{d}$ and right-handed
(s)leptons.

In the class of models that we consider, the inflaton is assumed to be
in a sector which is coupled to ordinary matter by interactions of
gravitational strength only.  In this case, the inflaton decay always
occurs in the perturbative regime and we need not worry about
parametric-resonance decay effects \cite{kls}.  The inflaton decay rate is
${\Gamma}_{d}
\sim
{{m}^{3} \over {{M}_{Pl}}^{2}}$, where $m$ is the inflaton mass and $m
\leq {10}^{13}$~GeV from the COBE data on the CMBR anisotropy \cite{cdo2}.
Efficient inflaton decay occurs at the time when $H \simeq
{\Gamma}_{d}$ and the effective reheat temperature at that time will be
${T}_{R} \sim {({\Gamma}_{d}{M}_{Pl})}^{{1 \over 2}}$.  For $m \sim {10}^{13}$ GeV
we
get ${T}_{R} \sim {10}^{10}$~GeV, which is in the allowed range to avoid
the gravitino problem \cite{ekn}.

The crucial point to note is that, although inflaton decay effectively
completes much later than the start of its oscillation, nonetheless
decay occurs throughout this period.  In fact, a dilute plasma with
temperature $T {\ \lower-1.2pt\vbox{\hbox{\rlap{$<$}\lower5pt\vbox{\hbox{$\sim$}}}}\ } {(H {\Gamma}_{d}{{M}_{Pl}}^{2})}^{{1 \over 4}} $
(assuming instant thermalization: we address thermalization below) is
present from the first several oscillations, until the effective
completion of the inflaton decay \cite{kt}.  It is easily seen that
it has the highest instantaneous temperature at the earliest time,
which can reach $T \leq {10}^{13}$ GeV.  This plasma,
however, carries a relatively small fraction of the cosmic energy
density, with the bulk still in inflaton oscillations. The dilution
of relics produced from this plasma by the entropy
release from the subsequent decay of the bulk of the inflaton energy is
the reason that it does not lead to gravitino overproduction.  It is
important to note that the energy density in the plasma may be comparable to the energy density stored in the condensate along
a flat direction.  As
a result, the thermal effects from the plasma may affect the dynamics of
flat direction evolution which, as we see below, occurs in many
cases.

All fields with mass less than $T$, and gauge interactions with the
plasma particles, can reach thermal equilibrium with the plasma.  Those
fields which are coupled to the flat direction have generically large
masses in the presence of its vev, and might not be excited
thermally.  These include the $\phi$ fields which are gauge-coupled to
$\alpha$
and
have a mass $g  \alpha$ (up to numerical factors of $O(1)$) and many
of the $\chi$ fields which have
superpotential
couplings to $\alpha$, and hence have a mass $h \alpha$ (also up to
numerical factors of $O(1)$).  For
$g \alpha > T$ or $h \alpha > T$, the former or the latter are not in
thermal equilibrium, respectively.  We recall that, in the presence of
Hubble-induced soft-breaking terms, the minimum of the potential for the
flat direction determines that $\alpha \sim {( H {M}^{n-3})}^{{1
\over n-2}}$, and the plasma temperature is $T \sim {(H
{\Gamma}_{d}{{M}_{Pl}}^{2})}^{{1 \over 4}} $.  So a field with a coupling
$h$ to the flat direction can be in thermal equilibrium provided that  $h
\alpha \leq T$, which implies that

\beq
 {H}^{6-n} \leq {{{\Gamma}_{d}}^{n-2} \over
{h}^{4 (n-2)}} {{{M}_{Pl}}^{2(n-3)} \over {M}^{4(n-3)}}
\eeq
and similarly for the gauge coupling $g$.

The back-reaction effect of the plasma of quanta of this field will then
induce a mass-squared $+{h}^{2}{T}^{2}$ for the flat direction to which
it is coupled.  If this exceeds the negative Hubble-induced mass-squared
$-{H}^{2}$, the flat direction starts its oscillation.
This happens for $ hT \geq H$, i.e., for

\beq
{H}^{3} \leq {h}^{4} {\Gamma}_{d}{{M}_{Pl}}^{2}
\eeq
and similarly for back-reaction from plasma fields with gauge coupling $g$
to
the flat direction.
Therefore, a flat direction will start its oscillations if both of the
above conditions are satisfied simultaneously.  We note that the
finite-temperature effects of the plasma can lead to a much earlier
oscillatory regime for the flat direction, i.e., when $H \gg {m}_{{3
\over 2}}$.

It is clear that, in order for a plasma of the quanta of a
field to be produced, the coupling of that field to a flat
direction should not be so large that its induced mass prevents its
thermal excitation.  On the other hand, in order for its thermal plasma
to have a significant reaction back on the flat direction, its
coupling to the flat direction should not be so small that the thermal
mass-squared induced for the flat direction will be 
smaller than the Hubble-induced
contribution.  Therefore, to have significant thermal effects, we need
couplings
of
intermediate strength in order to have both conditions simultaneously
satisfied.  For
the fields $\phi$ which have $D$-term couplings of gauge strength $g$ to
flat directions, this is usually not the case: As will be seen shortly, in
most
cases their
couplings are too large to satisfy the equilibrium condition.  
For the fields $\chi$ which have $F$-term 
couplings of Yukawa strength $h$ to the
flat direction, the existence of significant thermal effects depends on
the value of $h$, as well as on the initial value of the flat-direction
vev $\alpha$, which in turn depends on the mass scale and the dimension of
the higher-dimensional operator which lifts the flatness.

To organize our discusion then, we first assess the typical
values of $h$ to be expected for couplings to the flat directions. For
these typical values, we then estimate the importance of thermal
effects on vev's determined by higher-dimensional operators ranging
over the various different dimensions that can
lift the flat direction, for both the
case of lifting by the GUT scale: $O(10^{16})~{\rm GeV}$, and by the
Planck
scale:
$O(10^{19})~{\rm GeV}$.

We now list the Yukawa couplings of the MSSM.  For low $tan\beta$,
the ratio of ${H}^{u}$ and ${H}^{d}$ vev's, we have

\beq
\matrix{
{{h}^{u}}_{1} \sim {10}^{-4} & {{h}^{d}}_{1} \sim {10}^{-5} &
{{h}^{l}}_{1} \sim {10}^{-6} \cr
{{h}^{u}}_{2} \sim {10}^{-2} & {{h}^{d}}_{2} \sim {10}^{-3} &
{{h}^{l}}_{2} \sim {10}^{-3} \cr
{{h}^{u}}_{3} \sim 1 & {{h}^{d}}_{3} \sim {10}^{-2} & {{h}^{l}}_{3}
\sim {10}^{-2} \cr
}
\eeq
whilst 
the ${h}^{u}$'s and ${h}^{d}$'s tend to be more
similar for high $tan\beta$.
The only Yukawa couplings which are significantly different from
$O({10}^{-2})$ are ${{h}^{u}}_{1}$, ${{h}^{d}}_{1}$,
${{h}^{l}}_{1}$, and ${{h}^{u}}_{3}$. Only flat directions
which include only the left- and right-handed up squark, the left- and
right-handed down squark, the left- and right-handed selectron and
the left-handed sneutrino will have an $h$ significantly less than
$O({10}^{-2})$.

For low $tan\beta$, any flat direction which includes right-handed
top squarks has a Yukawa coupling of $O(1)$ to some $\chi$'s, too large for
those $\chi$'s to
be in thermal equilibrium, given the expected range of flat direction
vev's $\alpha$.  The left-handed squarks are coupled to both
${H}^{u}$ and ${H}^{d}$, so any flat direction which includes a
left-handed top squark has a Yukawa coupling of order ${10}^{-2}$ to
${H}^{d}$ as well.  For high $tan\beta$, any flat direction which
includes the left- or right-handed top or bottom squarks has a Yukawa
coupling of order 1 since the top and  bottom Yukawas are of the same order.
In general, any flat direction which consists only of the
above-mentioned scalars has a Yukawa coupling of order 1 to some $\chi$'s
and/or a Yukawa coupling significantly less than $O({10}^{-2})$ to other
$\chi$'s.

Among all flat directions which are not lifted by the renormalizable
superpotential terms, there is only one which allows such a flavor
choice: $uude$ with one $u$ in the third generation and all other
scalars in the first generation (i.e., $tude$).  This exceptional flat
direction
still
has a coupling of $O({10}^{-4})$ to some $\chi$ fields, since it includes
the right-handed up squark.  Taking into account all flavor choices for all
flat directions which are not lifted at the renormalizable superpotential
level, we can use $h \simeq {10}^{-2}$ for the coupling of a
generic flat directions to $\chi$ fields.  For the above-mentioned
exceptional case we shall use $h \simeq {10}^{-4}$.

So, for our discussion of the dynamics of flat direction oscillations we
will
consider three representative cases. We will analyze the dynamics when inflaton
decay plasmons are coupled to the flat direction by: gauge couplings with
coupling $g \simeq 10^{-1}$,  generic Yukawa couplings of order $h \simeq
10^{-2}$, or suppressed Yukawa couplings of order $h \simeq 10^{-4}$.
Consideration of these cases should allow us to explore the generic range
of physical
effects  that arise in flat direction oscillations, from a plasma of inflaton
decay products.



We now  undertake a detailed analysis to determine in which cases a plasma of
inflaton decay products can be produced, and can initiate the flat-direction
oscillations by the reaction they induce on the flat direction.  Whether this
occurs or not depends on the vev of the flat direction, and the strength of the
coupling of the plasma quanta to the flat direction. The initial vev of  the
flat direction is set by both the underlying scale of the physics of the
higher-dimensional operators that lift the flat direction, and, for a
given flat
direction, by the dimension of the gauge-invariant operator of lowest dimension
which can be induced by the underlying dynamics to lift the flat direction.

In order  to categorize systematically the various cases which arise, we
organize them as follows. First, we divide them into two cases, depending
on whether the underlying scale of the new physics responsible for the
higher-dimensional operators which lift the flat direction and stabilize the
vev at the end of inflation are GUT-scale: $O(10^{16})$~GeV, or
Planck-scale: $O(10^{19})$~GeV.  Each of these cases is subdivided
according
to whether the coupling between the flat direction and the inflaton decay
products is of gauge strength ($g \simeq10^{-1}$), standard superpotential
Yukawa strength ($h \simeq {10}^{-2}$), or exceptional suppressed Yukawa
strength ($h \simeq {10}^{-4}$). As noted above, this covers the generic
range of
couplings exhibited by fields in flat directions in the supersymmetric standard
model.  Finally, each of these cases is subdivided  and tabulated
according to
the dimension of the operator that stabilizes the flat-direction vev,
setting
(given the possibilities listed above for 
the underlying scale of the new physics responsible
for the operators) the initial vev of the flat direction. These
higher-dimensional operators are listed by the order of the monomial in the
superfields which appears in the superpotential and is responsible for the
operator.  We tabulate against the order of the higher-dimensional
superpotential term the following quantities (in Planck units
\footnote{From now on, we 
express some dimensionful quantities in Planck units.}):  the Hubble
constant  $H$, the temperature $T$, 
and the value of the flat-direction vev $\alpha$ 
at the onset of oscillations, as well as the 
combination ${h {T}^{2} \over H \alpha}$ (${g {T}^{2}
\over H \alpha}$ for the case of the gauge coupling) which will be useful
when we
discuss the produced baryon asymmetry in the next section.  We also explain
the reasons for the values of the entries appearing, in the light of the
two necessary conditions
introduced above for inducing the flat-direction oscillations by plasma
effects, i.e., that on the one hand the mass of the plasmon induced by
the coupling to the flat direction
is small enough that it can be populated in the thermal 
bath from inflaton decay, and, on
the other hand, that the coupling is large enough for back-reaction
effects from the
plasma to lift the flat direction sufficiently to start oscillation
despite the effects of the Hubble-induced mass.

First, let us consider the case that the scale of the new physics that
induces
operators that stabilize the flat direction is of order the GUT scale:
$O(10^{16})$~GeV.  We then subdivide this case according to 
the strength of
the coupling of the inflaton decay products to the flat direction.
To start, we consider the gauge-coupled case with
$g={10}^{-1}$. In
this case, it is only for initial flat-direction vev's fixed by either
quartic
or quintic higher-dimensional terms in the superpotential that the plasma
effects can accelerate the onset of flat direction oscillation, with the
results shown in the following Table.
Physically, for superpotential monomials of sixth order or higher, the initial
flat-direction vev is
sufficiently large that the mass generated by its gauge coupling to the
prospective
inflaton decay products is large enough to prevent them from
being kinematically accessible for thermal excitation.  In the case of a quintic
superpotential monomial this is also initially the case, and it is only after
Hubble expansion has reduced $\alpha$, and hence the
induced plasmon mass, that thermalized products of inflaton decay can
back-react
to induce flat-direction oscillation. However, this only occurs for $H <
10^{-16}$,
by which time the low-energy soft supersymmetry breaking has already
initiated flat-direction oscillation.

\begin{center}
{\it GUT scale $M = 10^{16}$ GeV, gauge coupling $g = 10^{-1}$}\\
{~~}\\
\begin {math}
\begin{tabular}{|c|c|c|c|c|} \hline
 & $H$ & $T$ & $\alpha$ & ${g {T}^{2} \over H \alpha}$ \\ \hline
$n=4$ & ${10}^{-8}$ & ${10}^{-{13 \over 2}}$ & ${10}^{-{11 \over 2}}$ &
${10}^{-{1 \over 2}}$ \\ \hline
$n=5$ & ${10}^{-18}$ & ${10}^{-{9}}$ & ${10}^{-{8}}$ &
${10}^{{7}}$ \\ \hline
\end{tabular}
\end{math}
\end{center}


For the GUT case $M={10}^{16}$ GeV with generic Yukawa coupling 
$h={10}^{-2}$, we have the results shown in the following Table for
lifting of the flat direction by monomials of
the orders listed.
In the cases that the order of the monomial is four or five we have no
difficulty satisfying the condition that $h \alpha \leq T$, so that they are
(thermally) populated in the inflaton decay plasma.  For monomials of order
six, seven or eight, the induced mass of the prospective plasmon is, in
fact, of the same
order or slightly larger than the instantaneous effective temperature.  So
thermally they are present, albeit now with some Boltzman suppression.
Moreover,
we also note that these induced masses are less than the mass of the decaying
inflaton, and so they will be produced in the cascade of inflaton decay
products, though, as noted above, after complete thermalization they will
be
subject to some Boltzmann suppression.  
In all cases the value of the Hubble constant at the onset of oscillation
will be determined by the second condition 
($hT \geq H$), which requires that the
back-reaction-induced mass overcome the Hubble-induced 
mass to initiate oscillation.  
By comparing the results of the Tables for $g = {10}^{-1}$ 
and $h = {10}^{-2}$, we note that for a general 
flat direction with $h = {10}^{-2}$ which is lifted at 
the $n = 4$ superpotential level, 
the values at the onset of oscillations should be taken from the
gauge analysis.
The reason is that, in this case, the back-reaction of the inflaton 
decay products which have gauge coupling to the flat direction 
act at an earlier time than the back-reaction of those 
decay products which have Yukawa couplings to it.

\begin{center}
{\it GUT scale $M = 10^{16}$ GeV, standard Yukawa coupling $h =
10^{-2}$}\\
{~~}\\
\begin {math}
\begin{tabular}{|c|c|c|c|c|} \hline
 & $H$ & $T$ & $\alpha$ & ${h {T}^{2} \over H \alpha}$ \\ \hline
$n=4$ & ${10}^{-{26 \over 3}}$ & ${10}^{-{20 \over 3}}$ & ${10}^{-{35 \over 6}}$ &
${10}^{-{2 \over 3}}$ \\ \hline
$n=5$ & ${10}^{-{26 \over 3}}$ & ${10}^{-{20 \over 3}}$ & ${10}^{-{44 \over
9}}$ & ${10}^{-{16 \over 9}}$ \\ \hline
$n=6$ & ${10}^{-9}$ & ${10}^{-{27 \over 4}}$ & ${10}^{-{64 \over 15}}$ &
${10}^{-2}$ \\ \hline
$n=7$ & ${10}^{-{28 \over 3}}$ & ${10}^{-{41 \over 6}}$ & ${10}^{-{64 \over
15}}$ & ${10}^{-{31 \over 15}}$ \\ \hline
$n=8$ & ${10}^{-{34 \over 3}}$ & ${10}^{-{22 \over 3}}$ & ${10}^{-{79 \over
18}}$ & ${10}^{-{17 \over 18}}$ \\ \hline
\end{tabular}
\end{math}
\end{center}


For $M={10}^{16}$ GeV, $h={10}^{-4}$, as a function of the order of the
superpotential monomial lifting the flat direction we have the
results shown in the next Table.
For  these cases, the flat-direction-induced
mass is always less
than the instantaneous temperature, due to the weak coupling of the flat
direction to the plasmons.
The only non-trivial condition now is the second  one ($hT \geq H$), which
determines how long one must wait before the Hubble-induced mass is
sufficiently
reduced that the back-reaction-induced flat-direction mass can overcome it
to
initiate oscillation. This fixes the value of $H$ at the onset of
oscillation. Comparing the results of the Tables for $g = {10}^{-1}$ and
$h =
{10}^{-4}$, we note that for an exceptional flat direction
with $h = {10}^{-4}$ which is lifted at
the $n=4$ superpotential level, the values at the onset of 
oscillations should also be taken from the gauge analysis.

\newpage
\begin{center}
{\it GUT scale $M = 10^{16}$ GeV, exceptional Yukawa coupling $h =
10^{-4}$}\\
{~~}\\
\begin {math}
\begin{tabular}{|c|c|c|c|c|} \hline
 & $H$ & $T$ & $\alpha$ & ${h {T}^{2} \over H \alpha}$ \\ \hline
$n=4$ & ${10}^{-{34 \over 3}}$ & ${10}^{-{22 \over 3}}$ & 
${10}^{-{43 \over 6}}$ & ${10}^{-{1 \over 2}}$ \\ \hline
$n=5$ & ${10}^{-{34 \over 3}}$ & ${10}^{-{22 \over 3}}$ & ${10}^{-{52 \over
9}}$ & ${10}^{-{14 \over 9}}$ \\ \hline
$n=6$ & ${10}^{-{34 \over 3}}$ & ${10}^{-{22 \over 3}}$ & ${10}^{-{61 \over
12}}$ & ${10}^{-{9 \over 4}}$ \\ \hline
$n=7$ & ${10}^{-{34 \over 3}}$ & ${10}^{-{22 \over 3}}$ & ${10}^{-{14 \over
3}}$ & ${10}^{-{8 \over 3}}$ \\ \hline
$n=8$ & ${10}^{-{34 \over 3}}$ & ${10}^{-{22 \over 3}}$ & ${10}^{-{79 \over
18}}$ & ${10}^{-{53 \over 18}}$ \\ \hline
$n=9$ & ${10}^{-{34 \over 3}}$ & ${10}^{-{22 \over 3}}$ & ${10}^{-{88 \over
21}}$ & ${10}^{-{22 \over 7}}$ \\ \hline
\end{tabular}
\end{math}
\end{center}




We now turn to the case that the underlying scale of new physics
responsible for generating the higher-dimensional operators is the Planck
scale. This means that the values of the flat direction vev's after inflation
will be larger, raising the mass of prospective plasmons to which they couple,
and making it harder to satisfy the constraint that these putative
plasmons be generated thermally, or even be kinematically accessible to
inflaton decay.

For $M={10}^{19}$ GeV, $g={10}^{-1}$, we have significant effects only for flat
directions lifted by superpotential terms arising from quartic or quintic
monomials. In all other cases ($n\geq6$) the flat direction vev is so large
that quanta gauge-coupled to it receive sufficiently large masses that
they can
not be thermally populated at the instantaneous temperature of the inflaton
decay products. For the two non-trivial cases we have the results
shown in the following Table.
Only in the $n=4$ case can we produce thermally a number
of plasma quanta sufficient to induce enough mass for the
flat direction to initiate its oscillation at an earlier time.  In the
$n=5$ case,
back-reaction from the plasma of  inflaton decay products only manages to
induce
flat-direction oscillation after  $H \ll 10^{-18}$, by which time the low-energy
soft supersymmetry breaking has already acted to start the oscillation and 
also the inflaton decay has been completed.

\begin{center}
{\it Planck scale $M = 10^{19}$ GeV, gauge coupling $g = 10^{-1}$}\\
{~~}\\
\begin {math}
\begin{tabular}{|c|c|c|c|c|} \hline
 & $H$ & $T$ & $\alpha$ & ${g {T}^{2} \over H \alpha}$ \\ \hline
$n=4$ & ${10}^{-14}$ & ${10}^{-{8}}$ & ${10}^{-{7}}$ &
${10}^{{4}}$ \\ \hline
$n=5$ & ${10}^{-42}$ & ${10}^{-{15}}$ & ${10}^{-{14}}$ &
${10}^{{25}}$ \\ \hline
\end{tabular}
\end{math}
\end{center}


For $M={10}^{19}$ GeV and $h={10}^{-2}$, we again have a case where flat
directions
lifted by superpotential monomials of order six or higher result in such a
large flat-direction vev that  quanta coupled to it receive too large a
mass
for them to be thermally excited in the plasma of inflaton decay products. For
the cases of quartic or quintic superpotential monomials, we have the
results shown in the following Table.
We again find that only in the $n=4$ case can thermal
effects
actually induce sufficient mass for the flat direction to
initiate oscillation  earlier.  In the $n=5$ case, back-reaction from the
plasma
of  inflaton decay products only manages to induce flat-direction
oscillation
after  the low-energy soft supersymmetry breaking has already done
so, and the inflaton decay has been completed.  By comparing the results
of the Tables for $g = {10}^{-1}$ and $h = {10}^{-2}$, we note that, for a
generic flat direction, i.e., one with $h = {10}^{-2}$, the initial values
at the onset of oscillations should be taken from the latter.  
The reason is that, in this case, the back-reaction of the inflaton decay
products which have Yukawa couplings to the flat direction act at an
earlier
time than the back-reaction of those decay products with a gauge coupling
to it.

\begin{center}
{\it Planck scale $M = 10^{19}$ GeV, standard Yukawa coupling $h =
10^{-2}$}\\
{~~}\\
\begin {math}
\begin{tabular}{|c|c|c|c|c|} \hline
 & $H$ & $T$ & $\alpha$ & ${h {T}^{2} \over H \alpha}$ \\ \hline
$n=4$ & ${10}^{-10}$ & ${10}^{-{7}}$ & ${10}^{-{5}}$ &
${10}^{-{1}}$ \\ \hline
$n=5$ & ${10}^{-30}$ & ${10}^{-{12}}$ & ${10}^{-{6}}$ &
${10}^{{14}}$ \\ \hline
\end{tabular}
\end{math}
\end{center}


Finally, in the case $M={10}^{19}$ GeV, $h={10}^{-4}$, for flat directions
lifted by monomials
higher than sixth order the resulting flat-direction vevs are sufficiently
large that quanta coupled to it with this coupling are too massive to be
excited at the instantaneous temperature of the inflaton decay products. So the
nontrivial cases are those in the following Table.
For $n=6$, we marginally satisfy the requirement that $hT \simeq \alpha$,
necessary for thermal production of quanta coupled to the flat
direction, while for $n=4$ and $n=5$ we do so comfortably. The second
condition, that  $hT \geq H$ for effective back-reaction, then serves to
determine
the value of $H$ at the onset of the thermally-induced oscillation.  By
comparing the results of the Tables for $g = {10}^{-1}$ and $h =
{10}^{-4}$,
we note that when the exceptional flat direction with $h =
{10}^{-4}$ is lifted at the $n = 4$
superpotential level, the values at the onset of 
oscillations should be taken from the latter.  
This is because the back-reaction of the inflaton 
decay products with Yukawa coupling to the flat direction 
act at an earlier time than the back-reaction of those 
decay products with gauge couplings to it.

\begin{center}
{\it Planck scale $M = 10^{19}$ GeV, exceptional Yukawa coupling $h =
10^{-4}$}\\
{~~}\\
\begin {math}
\begin{tabular}{|c|c|c|c|c|} \hline
 & $H$ & $T$ & $\alpha$ & ${h {T}^{2} \over H \alpha}$ \\ \hline
$n=4$ & ${10}^{-{34 \over 3}}$ & ${10}^{-{22 \over 3}}$ & ${10}^{-{17 \over
3}}$ & ${10}^{-{5 \over 3}}$ \\ \hline
$n=5$ & ${10}^{-{34 \over 3}}$ & ${10}^{-{22 \over 3}}$ & ${10}^{-{34 \over
9}}$ & ${10}^{-{32 \over 9}}$ \\ \hline
$n=6$ & ${10}^{-{38 \over 3}}$ & ${10}^{-{23 \over 3}}$ & ${10}^{-{19 \over
6}}$ &  ${10}^{-{7 \over 2}}$ \\ \hline
\end{tabular}
\end{math}
\end{center}



We noted above that thermal effects from the plasma can be important up to $h
\alpha \simeq T$ or even somewhat higher.  For $\alpha$ less than this, 
they
change the
convexity of the effective potential in the $\alpha$ direction at much
earlier times, inducing the onset of flat-direction oscillations.
We should note that since $\alpha \sim { H }^{{1 \over
n-2}}$ and $ T \sim { H}^{{1 \over 4}}$, then $\alpha$ decreases at the
same rate as, or more slowly than, $T$ for $7 \leq n \leq 9$.  This means
that
if
$h \alpha \gg T$ right after the end of inflation, it will remain so for
later times as well.  Therefore, in the $7 \leq n \leq 9$
cases for  $M={10}^{19}$ GeV, the Hubble-induced negative mass-squared is
dominant and
$\alpha$ will not be lifted until $H \simeq {10}^{-16}$, if $h \alpha \gg
T$ at $H \simeq {10}^{-6}$.

In sum, we conclude that for $M={M}_{GUT}$, a general flat direction with $h
\simeq {10}^{-2}$ starts oscillating at $H \gg {10}^{-16}$ in
the $4\leq n\leq8$ cases.  For the exceptional one with $h \simeq
{10}^{-4}$ it is true in the $n=9$ case as well.  For $M={M}_{Planck}$ in
the denominator, only in the $n=4$ case do oscillations of a general flat
direction start at $H \gg {10}^{-16}$.  In the $5 \leq n
\leq 9$ cases, the flat direction is protected from thermal effects
because its large vev induces such a large mass for fields coupled
to it that they cannot be thermally excited in the plasma of inflaton
decay
products.
For the exceptional flat direction with $h \simeq {10}^{-4}$, this
protection is weaker because of the smaller Yukawa coupling to $\chi$
(which therefore are lighter and can be excited in thermal
equilibrium) and, as a result,
oscillations start at $H \gg {10}^{-16}$ in the $n=5,~6$ cases also.


We need to elaborate on the implicit assumption that the $\chi$'s
($\phi$'s) are effectively thermal upon
production.  In the model that we study, the inflaton decays in the
perturbative regime, and the decay products have a momentum less than,
or comparable to,
the inflaton mass $m\sim{10}^{-6}$.   The $\chi$'s ($\phi$'s) which are
produced in two-body decays have a momentum of order $m$~\footnote{The 
$\phi$'s
generically have larger $\alpha$-induced masses
than do the $\chi$'s, so their production may be delayed until
$\alpha$ Hubble-dilutes to a smaller value.}.
It can easily be seen that the temperature at which
oscillations start (assuming thermal equilibrium) is $\approx {10}^{-7}$
in all the above cases.  Since the momentum of produced particles is greater
than the the average thermal momentum, the dominant process to reach
equilibrium is through the decay of $\chi$'s ($\phi$'s) to other particles
with smaller momenta.
However, the momentum of $\chi$'s ($\phi$'s) is very close to the average
thermal momentum.  Since thermalization does not change the energy density
in the plasma, the number density of $\chi$'s ($\phi$'s) 
is also close to its thermal distribution.  
Therefore, the plasma-induced mass-squared ${h}^{2}{{n}_{\chi} \over
{E}_{\chi}}$ (${g}^{2}{{n}_{\phi} \over {E}_{\phi}}$) 
from $\chi$'s ($\phi$'s) is of the same order as  
${h}^{2}{T}^{2}$ (${g}^{2}{T}^{2}$). 

\section{Thermal $A$ Terms and Baryo/Leptogenesis}

Motion along the angular direction is required for the build-up of
a
baryon or lepton asymmetry.  This is possible if a torque is exerted on
$\alpha$ or, equivalently, if $\alpha$ is not in one of the discrete
minima along the angular direction, when it starts oscillating.  These
discrete minima are due to the $A$ term part of the potential.  Before the
start of oscillations, the Hubble-induced $A$ terms are dominant, and the
locations of the minima are determined by them.  During inflation,
$\alpha$ rolls down towards one of these minima and rapidly settles there.  After inflation it tracks that minimum and there is no motion
along the angular direction \cite{drt}.  What is necessary then is a non-adiabatic
change in the location of the minima, such that at the onset of
oscillations $\alpha$ is no longer in a minimum along the angular
direction.  In the absence of thermal effects, $\alpha$ would start its
angular motion (as well as its linear oscillations) at $H \simeq {m}_{{3
\over 2}}$.
This occurs as a result of uncorrelated phases of the $A$
terms induced by the Hubble expansion and
low-energy supersymmetry breaking.  At this time,
the latter takes over from the former, and $\alpha$ will in general no
longer be in
a minimum along the angular direction.  This will lead to the
generation of a baryon or lepton asymmetry if $\alpha$ carries a
non-zero number of either \cite{drt}.

As we have seen above, due to thermal effects, in many cases the flat
directions start oscillating at much larger $H$.  At this time the
Hubble-induced $A$ terms are still much larger than the low-energy ones
from hidden-sector supersymmetry breaking.  In order to have angular
motion for $\alpha$, another $A$ term of size comparable to the 
Hubble-induced one, but with uncorrelated phase, is required.  Since it is
finite-temperature effects from the plasma that produce a mass-squared
which dominates the Hubble-induced one, one might expect that the same
effects also produce an $A$ term which dominates the Hubble-induced
$A$ term.  This is the only new effect that could produce such an
$A$ term with uncorrelated phase, as the thermal plasma is the only
difference from the standard scenario.

The simplest such thermal $A$ terms arise at tree-level from cross terms
from the following two terms in superpotential

\beq
h \alpha \chi \chi + {\lambda}_{n} {{\alpha}^{n} \over n{M}^{n-3}}
\eeq
which results in the contribution
\beq
h {\lambda}_{n} {{{\chi}^{*}}^{2}{\alpha}^{n-1} \over {M}^{n-3}} + h. c.
\eeq
in the scalar potential.  In thermal equilibrium, $<{{\chi}^{*}}^{2}>$ can be
approximated by ${T}^{2}$
and therefore the thermal $A$ term is of order

\beq
h{\lambda}_{n} {{T}^{2}{\alpha}^{n-1} \over {M}^{n-3}}
\eeq
There is another thermal $A$ term that arises from one-loop diagrams with
gauginos and fermionic partners of $\alpha$. It
results in a contribution, in the thermal bath, of order:

\beq
{\lambda}_{n} {\left( {g T \over 4 \pi \alpha}\right)^2}  {{T
{\alpha}^{n}} \over {M}^{n-3}}
\eeq
We have checked that, for the parameter range of interest for this
process, this has the same order of magnitude as the
tree-level $A$ term.  In the following, we use the tree-level term
for our estimates.

The ratio of the thermal $A$ term to the Hubble-induced one is
${h
{T}^{2} \over H \alpha}$.  It is clear from the results summarized in the
Tables that, at the onset of $\alpha$ oscillations, the thermal
$A$ term is weaker than the Hubble-induced one in all cases.  
Therefore, at this time, the minimum along the angular
direction is slightly shifted, the curvature at  the minimum is
still determined by the Hubble-induced $A$ term, and the force in the
angular direction is of order 
$ {\lambda}_{n}{h{T}^{2}{\alpha}^{n-2} \over {M}^{n-3}}$.  
The ratio of the thermal $A$ term to the Hubble-induced one grows,
however, as ${h {T}^{2} \over H \alpha}$ increases in time.  
Therefore, in what follows we keep both $A$ terms in the equation of
motion of $\alpha$.

We consider the case where oscillations start because of the back-reaction
of the $\chi$ fields, as is the most common case. The masses
of
${\alpha}_{R}$ and ${\alpha}_{I}$ are then of order $hT$~\footnote{We
use the estimate $gT$ for the plasma-induced masses if oscillations start
because of the back-reaction of the $\phi$ fields.}.
The equation of motion for the flat direction is then:

\beq
\ddot{\alpha} + 3 H \dot{\alpha} + {h}^{2} {T}^{2} 
\alpha + (n-1) h {\lambda}_{n} {{T}^{2}{{\alpha}^{*}}^{n-2} 
\over {M}^{n-3}} + A {\lambda}_{n} {H {{\alpha}^{*}}^{n-1} 
\over {M}^{n-3}} + (n-1){{\lambda}_{n}}^{2} {{|\alpha|}^{2(n-2)} 
\over {M}^{2(n-3)}} 
\alpha \; = \; 0
\label{seveneight}
\eeq
At this time, the universe is matter-dominated, by the
oscillating inflaton field, and thus $H = {2\over 3t}$.  Also, ${T}^{2} =
{(H {\Gamma}_{d} {{M}_{Pl}}^{2})}^{{1 \over 2}} \sim {t}^{-{1 \over 2}}$ 
for $H \geq {10}^{-18}$. After re-scaling $
\alpha \rightarrow {({{H}_{osc} {M}^{n-3} \over 
{\lambda}_{n}} )}^{1 \over n-2} \alpha$ and 
$t \rightarrow {H}_{osc}t$, where ${H}_{osc}$ is the Hubble constant at
the
onset of oscillations, we get the following equations 
of motion for the real and imaginary components of $\alpha$:

\beq
\matrix{
\ddot{{\alpha}_{R}} + {2 \over t} \dot{{\alpha}_{R}} + 
a{{\alpha}_{R} \over {t}^{1 \over 2}} + 
b {{|\alpha|}^{n-2} \over {t}^{1 \over 2}} \cos{((n-2) \theta + \varphi)} + 
A {{|\alpha|}^{n-1} \over t} \cos{((n-1) \theta)} + 
(n-1){|\alpha|}^{2(n-2)} {\alpha}_{R} \; = \; 0 \cr
\cr
\cr
\ddot{{\alpha}_{I}} + {2 \over t} \dot{{\alpha}_{I}}  + 
a{{\alpha}_{I} \over {t}^{1 \over 2}} - 
b {{|\alpha|}^{n-2} \over {t}^{1 \over 2}} \sin{((n-2) \theta + \varphi)} - 
A {{|\alpha|}^{n-1} \over t} \sin{((n-1) \theta)} + 
(n-1){|\alpha|}^{2(n-2)} {\alpha}_{I} \; = \; 0 \cr  
}
\eeq
Here $a \equiv {{h}^{2}{{T}_{osc}}^{2} \over {{H}_{osc}}^{2}}$: ${T}_{osc}
=
{({H}_{osc} {\Gamma}_{d} {{M}_{Pl}}^{2})}^{1 \over 2}$ is the plasma 
temperature at the onset of oscillations, 
$b \equiv (n-1) a {\lambda}_{n} {{({{H}_{osc} {M}^{n-3} \over {\lambda}
_{n}})}^{n-3 \over n-2} \over {M}^{n-3}}$, 
and $\varphi = O(1)$ is the relative phase between the thermal
and Hubble-induced $A$ terms.

The first two terms and the superpotential term in each of these equations 
are the same as in the equations derived in~\cite{drt}, but there
are some important differences.  First of all, the flat-direction
mass-squared is not the (constant) low-energy value ${{m}_{3 \over
2}}^{2}$, but the thermal mass which is redshifted as ${t}^{-{1 \over 2}}$.  
Also, the Hubble-induced $A$ term with coefficient $H$ appears instead of 
the low-energy one with coefficient ${m}_{3 \over 2}$, which is negligible 
for $H \gg {10}^{-16}$.  This explains the ${1 \over t}$ factor in front 
of the Hubble-induced $A$ term.  Finally, there is another $A$ term, the
thermal one, which is also redshifted as ${t}^{-{1 \over 2}}$,
because of its ${T}^{2}$ dependence.

At the onset of oscillations, $t = {t}_{i} =  {2 \over 3}$ and $\alpha$ is
in one of the minima which are determined by the Hubble soft terms.  
Therefore, ${|\alpha|}_{i}$, which was ${({{H}_{osc} {M}^{n-3} \over
{\lambda}_{n}})}^{1 \over n-2}$ before re-scaling, is scaled to
${|\alpha|}_{i}=1$, and also ${\theta}_{i} = n \pi$.   
We have solved these equations numerically 
for ${\theta}_{i}=0$, $A=(n-1)$ and ${\lambda}_{n} =1$,
and calculated ${n}_{B} =
{\alpha}_{R}{\partial {{\alpha}_{I}}\over
 \partial t} - {\alpha}_{I}{\partial {{\alpha}_{R}}\over \partial I}$.  
We find that, among the cases listed in the above-mentioned Tables, only
for the following ones do we get an ${{n}_{B} \over s}$ of order ${10}^{-11}$ 
or larger, before any subsequent dilution after reheating.

\newpage
\begin{center}
{\it The value of ${{n}_{B} \over s}$ for flat directions which undergo plasma-induced oscillations}\\
{~~}\\
\begin{tabular}{|c|c|c|c|c|} \hline
 & \multicolumn{2}{c|}{$M = {10}^{16}$ GeV} & \multicolumn{2}{c|}{$M = {10}^{19}$ GeV} \\ \cline{2-5}
 & $h={10}^{-2}$ & $h={10}^{-4}$ & $h={10}^{-2}$ & $h={10}^{-4}$ \\ \hline
$n=4$ & $< {10}^{-11}$ & $< {10}^{-11}$ & $< {10}^{-11}$ & $3 \times {10}^{-11}$ \\ \hline
$n=5$ & $< {10}^{-11}$ & $3 \times {10}^{-11}$ & ${\rm no ~plasma ~effect}$ & $3 \times {10}^{-9}$ \\ \hline
$n=6$ & ${10}^{-11}$ & $4 \times {10}^{-11}$ & ${\rm no ~plasma ~effect}$ & ${10}^{-7}$ \\ \hline
$n=7$ & $4 \times {10}^{-11}$ & ${10}^{-10}$ & ${\rm no~ plasma ~effect}$ & ${\rm no ~plasma~ effect}$ \\ \hline
$n=8$ & ${10}^{-10}$ & $3 \times {10}^{-10}$ & ${\rm no ~plasma ~effect}$ & ${\rm no ~plasma~ effect}$ \\ \hline
$n=9$ & ${\rm no~ plasma ~effect}$& $5 \times {10}^{-10}$ & ${\rm no ~plasma ~effect}$ & ${\rm no ~plasma ~effect}$ \\ \hline
\end{tabular}
\end{center}

We see that, in some cases, ${{n}_{B} \over s}$ is near the observed value
of $ 5 \times {10}^{-10}$.  However, in the most general case, when the
standard model gauge group is the only symmetry group, 
these viable flat directions constitute only a small
subset of all flat directions. We also see that ${{n}_{B} \over s}$ is
larger for the exceptional flat directions, when $M = {10}^{19}$~GeV,
and for flat directions which are lifted by terms of higher order $n$.
This is easily understandable, as 
for larger $M$ and $n$, and for smaller $h$, plasma-induced 
oscillations start later and closer to the efficient reheat 
epoch $H = {10}^{-18}$.  Larger $M$ and $n$ lead to a larger vev for 
the flat direction and, therefore, the condition $h \alpha \leq T$
will be satisfied at a later time.  A smaller value of $h$, on the other
hand, implies that the condition $hT \geq H$ will be satisfied at a later
time.  Later oscillations mean less dilution of the generated 
lepton/baryon asymmetry by the plasma of inflaton decay 
products (we recall that $s \sim {T}^{3}$ is redshifted only as ${t}^{-{3
\over 4}}$ for $H \geq {10}^{-18}$).    

Now we comment how our results may be affected by changes in the
model-dependent constants involved in the calculations: the reheat temperature
${T}_{R}$ (or equivalently the inflaton decay rate), and the constant ${{C}_{I}
\over (n-1){\lambda}_{n}}$ which appears in the expression for the
flat-direction vev.  There are two concerns in this regard.  
First, whether the two
conditions for plasma-induced $\alpha$ oscillations still result in a
consistent value
for ${H}_{osc}$ which is greater than ${10}^{-16}$, and, secondly, what
is the corresponding
change in the estimated value for ${{n}_{B} \over s}$.  In
our calculations, we have used ${T}_{R} \simeq {10}^{-9}$ and ${{C}_{I}
\over
(n-1){\lambda}_{n}} \simeq 1$.  If we assume instead that ${T}_{R} \simeq
{10}^{-10}$
and ${{C}_{I} \over (n-1){\lambda}_{n}} \simeq {10}^{-1}$,
it turns out that for all cases except the marginal ones (the $n=6,7,8$
cases for $h={10}^{-2}$ and $M={10}^{16}$ GeV, and the $n=6$ case for
$h={10}^{-4}$
and $M={10}^{19}$~GeV), plasma effects still trigger the oscillations for
$H \geq
{10}^{-16}$, though at a somewhat smaller value of $H$.  Moreover, the
value of ${{n}_{B} \over s}$ remains the same within an order of
magnitude.
Therefore, the plasma-induced oscillations of the (non-marginal) flat
directions, and the resulting value of ${{n}_{B} \over s}$, are rather
insensitive to the exact order of magnitude of ${T}_{R}$ and ${{C}_{I} \over
(n-1){\lambda}_{n}}$, at least for our purposes.


\section{Evaporation of the Flat Direction}

Now let us find the time when the $\alpha$ condensates are knocked out of
the zero mode by the thermal bath.  For the evaporation to happen, it is
necessary that the thermal bath includes those particles 
which are coupled to $\alpha$.  
Then, two conditions should
be satisfied: first, the scattering rate of $\alpha$ off the thermal bath
must be sufficient for
equilibriation, and secondly, the energy density in the bath must be
greater than
that in the condensate.  The flat direction has couplings both of Yukawa
strength $h$ to $\chi$'s and of gauge strength $g$ to $\phi$'s.  
The conditions for thermal production of $\chi$'s and $\phi$'s 
are $h \alpha \leq T$ and $g \alpha \leq T$,
respectively.  Since $h < g$, the $\chi$'s will come to 
thermal equilibrium at an earlier time.  On the other hand, 
the scattering rate of $\alpha$ off the thermal $\chi$'s 
is ${\Gamma}_{scatt.} \sim {h}^{4}T$ while the rate for 
scattering of $\alpha$ off the 
thermal $\phi$'s is ${\Gamma}_{scatt.} \sim {g}^{4}T$.  
Therefore, $\chi$'s are produced earlier but in general have a 
smaller scattering rate.  The competition between the $\chi$'s 
and $\phi$'s, and between the ratio of the energy density of the 
flat direction to the energy density in the plasma will 
determine whether and how the flat direction evaporates. 

First we consider those flat directions which have plasma-induced
oscillations.  If oscillations start due to the back-reaction 
of $\chi$'s (which is the situation for most cases) 
${\Gamma}_{scatt.} \sim {h}^{4}T$.  
For a general flat direction with $h \simeq {10}^{-2}$, this is comparable
to $H$ at $H \simeq {10}^{-17}$, while for the exceptional 
flat direction with $h \simeq {10}^{-4}$ this occurs 
at a much smaller $H$.  However, after $\alpha$ starts its
oscillation, it is redshifted as ${t}^{-{7
\over 8}}$ while $T$ is redshifted as ${t}^{-{1 \over 4}}$.  
This implies that ${g
\alpha \over T}$ decreases rapidly and soon the $\phi$'s will be in 
thermal
equilibrium.  The rate for scattering of
$\alpha$ off thermal $\phi$'s is ${\Gamma}_{scatt.} 
\sim {g}^{4}T$ and ${\Gamma}_{scatt.} \geq H$ at
$H \leq {10}^{-12}$.  The energy density in the 
condensate at the onset of oscillations is
${h}^{2}{\alpha}^{2}{T}^{2} \leq {T}^{4}$ 
(recall that $h \alpha \leq T$ at this time).  
The ratio of the two energy densities is further redshifted as
${t}^{-{5 \over 4}}$ (for $H \geq {10}^{-18}$) 
which ensures the second necessary
condition for the evaporation of condensate, i.e., that the plasma energy
density is
dominant over the energy density in the condensate.  It can easily
be checked
that the condensate evaporates at $H \gg {10}^{-18}$, before the inflaton decay
is completed
\footnote{If $\alpha$ oscillations start due to the back-reaction of
$\phi$'s, from the beginning $g \alpha \leq T$ and ${\Gamma}_{scatt.} \sim
{g}^{4}T$.  Therefore, there is no need to wait for further redshift of
$\alpha$ and again the condensate evaporates at $H \gg {10}^{-18}$.}.  

In those cases in which the plasma effects do not lead to an early
oscillation of the flat direction, oscillations start at $H \simeq {10}^{-16}$,
when the low-energy supersymmdetry breaking takes over the 
Hubble-induced one.  It is important to find the time when the condensate will
evaporate in these cases too.  For such flat directions, the ratio
of the baryon
number density to the condensate density is of order one \cite{drt}.
Therefore, if the condensate dominates the energy density of the
universe before evaporation, the resulting ${{n}_{B} \over s}$ will
also be of order one.  Some regulating mechanism is then needed in
order to obtain the value for successful big bang nucleosynthesis: 
${{n}_{B}
\over s}\sim {10}^{-10}$
\cite{cgmo}.

Now consider a general flat direction with $h \simeq {10}^{-2}$.  As
we showed, in the $5 \leq n \leq 9$ cases for $M={10}^{19}$ GeV, and the $n=9$ case for $M={10}^{16}$ GeV, plasma effects are not
important and the flat direction starts oscillating at $H \simeq {10}^{-16}$.  By
$H \simeq {10}^{-18}$ the inflaton has efficiently decayed and $\alpha$ has
been redshifted by a factor of ${10}^{-2}$.  From then on, the universe is
radiation-dominated, so $\alpha \propto {t}^{-{3 \over 4}}$ and $T
\propto
{t}^{-{1 \over 2}}$.  Therefore, the energy density in the condensate is
redshifted as $ {t}^{-{3 \over 2}}$ whilst the energy density in
radiation is redshifted as $ {t}^{-2}$.  If the condensate does not
evaporate (or decay) until very late times, its energy density
dominates that of the radiation and universe will again be
matter-dominated.  At the beginning of oscillations, i.e., at $H \simeq
{10}^{-16}$, $\alpha$ has the largest
vev in the $n=9$ case for $M={10}^{19}$ GeV, which is 
$\alpha \simeq {10}^{-{16 \over 7}}$.  
At $H \simeq {10}^{-18}$ this is redshifted to $\alpha \simeq {10}^{-{30 \over
7}}$ which still leaves $h \alpha > T$, 
so plasmons with this Yukawa coupling to the flat 
direction cannot be produced.  However, since $\alpha$ redshifts more
rapidly than $T$, eventually $h \alpha$ becomes of order $T$, 
after a time such that

\beq
T\simeq {10}^{-{101 \over 7}},~\alpha \simeq
{10}^{-{87 \over 7}}
\eeq
It is easily seen that at this time the energy density in the
condensate and in the radiation are of the same order.  Moreover,
${\Gamma}_{scatt.}\sim {10}^{-8}T\gg H$ and the 
condensate evaporates promptly.  This case is marginal as the condensate
almost dominates the energy density of the universe at evaporation.

In the $5 \leq n \leq 8$ cases for $M = {10}^{19}$ GeV 
and the $n=9$ case for $M={10}^{16}$ GeV the vev is 
considerably smaller and the energy
density in radiation is even more dominant.  Therefore, a general flat
direction with $h \simeq {10}^{-2}$ will evaporate before dominating the
energy density of the universe.  We summarize the 
situation for a general flat direction with $h \simeq {10}^{-2}$, 
regarding both the early, i.e., plasma-induced, oscillation, 
and evaporation, in the following Table.

\begin{center}
{\it Viability of scenarios with generic Yukawa coupling $h = 10^{-2}$}\\
{~~}\\
\begin{tabular}{|c|c|c|c|c|} \hline
 & \multicolumn{2}{c|}{$M = {10}^{16}$ GeV} & \multicolumn{2}{c|}{$M = {10}^{19}$ GeV} \\ \cline{2-5}
 & Early Oscillation & Evaporation & Early Oscillation & Evaporation \\ \hline
$n=4$ & $\surd$ & $\surd$ & $\surd$ & $\surd$ \\ \hline
$n=5$ & $\surd$ & $\surd$ & & $\surd$ \\ \hline
$n=6$ & ${\rm marginal}$ & $\surd$ & & $\surd$ \\ \hline
$n=7$ & ${\rm marginal}$ & $\surd$ & & $\surd$ \\ \hline
$n=8$ & ${\rm marginal}$ & $\surd$ & & $\surd$ \\ \hline
$n=9$ &  & $\surd$ & & ${\rm marginal}$ \\ \hline
\end{tabular}
\end{center}

For the exceptional flat direction with $h \simeq {10}^{-4}$ the situation is
different.  Here plasma effects are not important 
in the $7 \leq n \leq 9$ cases for $M = {10}^{19}$ GeV.  
In the $n=9$ case the condition for thermal 
production of $\chi$'s, $h \alpha = T$ gives

\beq
T\simeq {10}^{-{73 \over 7}},~\alpha \simeq
{10}^{-{45 \over 7}}
\eeq
which means we do not need as much redshift to 
reduce $\alpha$, so $\chi$'s are produced earlier and at a higher temperature.
However, ${\Gamma}_{scatt.} \sim {10}^{-16}T$, which is much smaller than
$H$ at this time.  Therefore, the condensate
cannot evaporate by scattering off the $\chi$'s.  
It is easily seen that $hT > {m}_{{3
\over 2}} \simeq {10}^{-16}$ when $h \alpha = T$.  
This implies that the mass and energy density of the 
flat direction are $hT$ and ${h}^{2}{\alpha}^{2}{T}^{2}$, 
respectively, upon thermal production of $\chi$'s, and the energy density
in the flat direction and the thermal bath are comparable.  As long as $hT \geq
{m}_{{3 \over 2}}$, $\alpha$
and $T$ are both redshifted as ${t}^{-{1 \over 2}}$.  
During this interval ${\alpha \over T}$ remains
constant and the flat direction and plasma energy densities remain
comparable.  Later, when $T < {10}^{-12}$ we have $hT< {m}_{{3 \over
2}}$, and the energy
density in the condensate is ${{m}_{{3 \over 2}}}^{2}{\alpha}^{2}$ 
and begins to dominate the thermal energy density.  
At some point, $g \alpha < T$ and $\phi$'s can be
produced thermally.  The scattering rate of the condensate off the $\phi$'s is
${\Gamma}_{scatt.}\sim {10}^{-4}T$ which is clearly at equilibrium.  However,
the energy density
in the condensate is now overwhelmingly dominant and evaporation
does not occur.  For the $n=7,~8$ cases the situation is similar and the
condensate does not evaporate.  The summary for the exceptional flat
direction, regarding both the early, i.e., plasma-induced, oscillation,
and evaporation, is illustrated in the Table below.

\begin{center}
{\it Viability of scenarios with exceptional Yukawa coupling $h =
10^{-4}$}\\
{~~}\\
\begin{tabular}{|c|c|c|c|c|} \hline
 & \multicolumn{2}{c|}{$M = {10}^{16}$ GeV} & \multicolumn{2}{c|}{$M = {10}^{19}$ GeV} \\ \cline{2-5}
 & Early Oscillation & Evaporation & Early Oscillation & Evaporation \\ \hline
$n=4$ & $\surd$ & $\surd$ & $\surd$ & $\surd$ \\ \hline
$n=5$ & $\surd$  & $\surd$ & $\surd$ & $\surd$ \\ \hline
$n=6$ & $\surd$  & $\surd$ & ${\rm marginal}$ & $\surd$ \\ \hline
$n=7$ & $\surd$  & $\surd$ &  &  \\ \hline
$n=8$ & $\surd$  & $\surd$ &  &  \\ \hline
$n=9$ & $\surd$  & $\surd$ &  &  \\ \hline
\end{tabular}
\end{center}

In summary: a general flat direction, i.e., with $h \simeq {10}^{-2}$,
which does not have plasma-induced early oscillation, does not come to
dominate the energy density of the
universe (the $n=9$ case for $M={10}^{19}$ GeV is marginal).
For the exceptional flat direction, i.e., with $h \simeq {10}^{-4}$, the
situation is
different and it dominates the energy density of the universe before decay.


\section{Discussion}

We have found that all flat directions, except those which are lifted by
nonrenormalizable superpotential terms of high dimension and with a large mass
scale in the denominator, start oscillating at early times due to
plasma effects.  For a general flat direction with $h \simeq {10}^{-2}$ these
are the $n=4$ case for $M={10}^{19}$ GeV and the $4 \leq n \leq8$ cases
for $M={10}^{16}$ GeV (with the $6 \leq n \leq 8$ cases being marginal and
sensitive to model-dependent parameters).  For the exceptional flat
direction
with
$h \simeq {10}^{-4}$ these are the $4 \leq n \leq 6$ cases for $M={10}^{19}$
GeV (with the $n=6$ case being marginal and sensitive to model-dependent
parameters) and all $n$ for $M = {10}^{16}$ GeV.  In these cases it is difficult to achieve efficient baryon asymmetry generation by the oscillation of the condensate along the flat direction.  We showed that a general
flat direction, i.e., one with $h \simeq {10}^{-2}$, which is not lifted
by thermal
effects, still evaporates before dominating the energy density of the universe.  This
is not important for baryogenesis, however, and the resulting dilution
by the thermal bath can be used to regulate the ${{n}_{B} \over s}$
which is initially of order one. On the other hand,
the exceptional flat direction, i.e., one with $h
\simeq {10}^{-4}$, which is not lifted by plasma effects, 
dominates the energy density of the universe before its decay.

For models with supersymmetry breaking via low-energy gauge mediation,
on the other hand, the evaporation of the condensate has yet another
implication.  In such models there is a candidate for cold dark matter, the
so called Q-ball \cite{ks}.  In order to have stable Q-balls as dark matter
candidates, some flat directions must dominate the energy density of the
universe.  This means that any flat direction which is evaporated by
the thermal bath cannot be used to form a Q-ball.

Now the question is which flat directions are lifted by $n> 4$
terms.  A look at \cite{gkm} reveals that only 18 out of 295
directions which are $D$- and $F$-flat at the renormalizable level in
the
MSSM are not lifted at the $n=4$ level.  Even a smaller subset of only
2 flat directions are not lifted at the $n=6$ level.  If
nonrenormalizable terms with $n=4$ and $n=5$ are not forbidden by
imposing other symmetries, only a very few flat directions in the MSSM
can be used for baryogenesis and even fewer for Q-ball formation
(regardless of the mass scale in the denominator or Yukawa couplings of
these flat directions).  This is if all higher-order terms which respect
gauge
symmetry exist in the superpotential.  With other symmetries (discrete or
continuous) imposed on the model, a specific flat direction will, 
in general, be lifted at a
higher level.  The initial vev of $\alpha$ can then be larger and $\chi$ and
$\phi$ quanta may not be produced thermally, and the standard treatment of the
Affleck-Dine baryogenesis may be valid.  Model-dependent analysis is
needed to identify at which level a given flat direction is actually
lifted, in a given model.

Finally, an interesting possibility is the parametric-resonance decay of a
supersymmetric flat direction to the fields $\phi$ to 
which it is gauge-coupled.  The occurence and implications of a potential
parametric resonance are more pronounced for those flat 
directions which start their oscillations at $H \simeq {10}^{-16}$, 
as in the standard scenario.  They have an incredibly 
large $q = {({g \alpha \over 2{m}_{3 \over 2}})}^{2}$~\footnote{For those
flat directions which have plasma-induced oscillations,
${m}_{3 \over 2}$ is replaced by $hT$ or $gT$, leading 
to a considerably smaller $q$.} which could be as large 
as $O({10}^{20})$ (the parameter $q$ determines the strength of 
resonance \cite{kls}).  Explosive resonance decay could also 
prevent these flat directions from dominating the energy density 
of the universe.  However, the situation is too complicated to allow 
simple estimates based on the results of parametric-resonance 
decay of a real scalar field.  First of all, the renormalizable 
part of the scalar potential (including the $D$-term part which 
is responsible for parametric-resonance decay to $\phi$'s) is 
fully known and very complicated.  Moreover, the flat direction itself
is a complex scalar field.  This may result in out-of-phase oscillations 
in the imaginary part of the flat direction, as well as in other 
scalar fields which are coupled to the same $\phi$, which can 
then substantially alter the outcome of simple parametric 
resonance \cite{acs1}.       

\section{Conclusion}

In conclusion, we have seen that many of the MSSM flat directions may
start their oscillations differently than in the standard scenario, 
where the low-energy supersymmetry breaking determines the onset of 
oscillations.  The two key ingredients for such a
different behaviour are: superpotential Yukawa couplings 
of the flat directions to other fields, and the thermal 
plasma from partial inflaton decay, whose instantaneous 
temperature is higher than the reheat temperature.  Together, 
these lead to an earlier
start of the oscillations.  On the one hand, the masses of
those fields 
which are coupled to 
the flat direction that are induced by the flat-direction vev
are then small enough to be kinematically accessible 
to inflaton decay and, on the other hand, induce large enough thermal 
masses for flat directions from the back-reaction of those fields to 
overcome the negative Hubble-induced mass-squared of the flat 
directions.  Subsequently, thermal masses and $A$ terms may be 
responsible for baryo/leptogenesis, but typically result 
in an insufficient baryon/lepton asymmetry of the universe.  
The oscillations are also terminated earlier, due to evaporation 
of the flat direction through its interactions with the thermal 
plasma.  It was also shown that even for many flat directions 
whose oscillations are not initiated by plasma effects, 
these effects cause them to evaporate before dominating 
the energy density of the universe.

{~~}\\
\noindent{ {\bf Acknowledgements} } \\
{~~}\\
\noindent 
The work of RA and BAC was supported in part by the Natural Sciences and
Engineering Research Council of Canada, and they would also like to thank
the CERN Theory Division for kind hospitality during part of this
research.

\end{document}